\documentclass[10pt,conference,table,xcdraw,dvipsnames]{IEEEtran}
\usepackage[utf8]{inputenc}
\usepackage{amsmath} 
\usepackage{amsthm} 
\usepackage{amssymb} 
\usepackage{graphicx} 
\usepackage{multicol} 
\usepackage{sparklines} 
\usepackage{subcaption}
\usepackage{hyperref}
\usepackage{cleveref}
\usepackage{comment}
\usepackage{booktabs}
\usepackage{adjustbox}
\usepackage{url}
\usepackage{fontawesome}
\usepackage{orcidlink}

\let\svthefootnote\thefootnote
\newcommand\blankfootnote[1]{%
  \let\thefootnote\relax\footnotetext{#1}%
  \let\thefootnote\svthefootnote%
}

\newcommand{\MyBox}[1]{\vspace{3mm}\noindent\framebox[\columnwidth][c]{\parbox[b]{0.95\columnwidth}{ #1 }}\vspace{3mm}}

\title{Seeking Enlightenment: Incorporating Evidence-Based Practice Techniques in a Research Software Engineering Team}
\date{October 2023}

\author{\IEEEauthorblockN{Reed Milewicz\orcidlink{0000-0002-1701-0008},
Jonathan Bisila\orcidlink{0009-0000-9618-5836}, Miranda Mundt\orcidlink{0000-0002-5283-2138}, and
Joshua B. Teves\orcidlink{0000-0002-7767-0067}}
\IEEEauthorblockA{Sandia National Laboratories\\1515 Eubank Blvd SE\\Albuquerque, New Mexico 87123\\
	\{rmilewi,jbisila,mmundt,jbteves\}@sandia.gov}}

\begin{document}

\maketitle

\begin{abstract}
    Evidence-based practice (EBP) in software engineering aims to improve decision-making in software development by complementing practitioners' professional judgment with high-quality evidence from research. We believe the use of EBP techniques may be helpful for research software engineers (RSEs) in their work to bring software engineering best practices to scientific software development. In this study, we present an experience report on the use of a particular EBP technique, rapid reviews, within an RSE team at Sandia National Laboratories, and present practical recommendations for how to address barriers to EBP adoption within the RSE community.
\end{abstract}


\section{Introduction}
\label{sec:introduction}

Modern science and engineering relies heavily on collaboratively-developed ecosystems of software that are both complex and constantly evolving.\blankfootnote{Sandia National Laboratories is a multimission laboratory managed and operated by National Technology \& Engineering Solutions of Sandia, LLC, a wholly owned subsidiary of Honeywell International Inc., for the U.S. Department of Energy's National Nuclear Security Administration under contract DE-NA0003525. SAND2023-06689C. 

Presented at USRSE2023, Chicago, IL, October 16-18, 2023.} Research software engineers (RSEs) are instrumental in the design, development, and use of that software. Since the term RSE was coined over ten years ago\cite{baxter2012research}, the RSE community has grown significantly, and we must consider how best to retain and cultivate that talent moving forward. In this work, we consider the use of \textbf{evidence-based practice} (EBP) as a potential strategy to help RSEs. The goal of EBP in software engineering is to integrate current best evidence from research with practical experience and human values to improve decision-making related to software development and maintenance\cite{dyba2005evidence}. 

We argue that the evidence-based paradigm could help RSEs by enabling career-long learning, furthering the professionalization of the field, and advancing better standards of practice---as in other disciplines where it has been successfully adopted. At present, while EBP techniques are popular among software engineering researchers, the use of EBP by software engineers is a relatively understudied and underdeveloped area of practice. Moreover, to the best of our knowledge, this work is the first to explore EBP within research software engineering specifically.

To lay a foundation for future work and dialogue, we present an experience report on the use of EBP techniques within an RSE team, the Department of Software Engineering and Research at Sandia National Laboratories\cite{milewicz2020department}\cite{willenbring2020department}. Starting in 2020, the first author, a software engineering researcher, began offering rapid reviews as a service to other members of the department, which are time-boxed literature reviews meant to quickly translate findings from research into actionable advice for practitioners\cite{cartaxo2018role}. While this was done for practical reason of supporting the high-quality work of our team, it was also an opportunity to gather data to inform the use of EBP techniques among the RSE community more broadly. Our key research questions were:

\begin{itemize}
	\item \textbf{RQ1}: What are the strengths and limitations of rapid reviews as an EBP technique in RSE contexts?
	\item \textbf{RQ2}: What are the challenges for RSEs to adopt EBP techniques more generally? What strategies would help address those challenges?
\end{itemize}

In addition to introducing readers to EBP concepts and techniques, our primary contribution is an in-depth analysis of a selection of carefully documented rapid reviews from two perspectives: (1) a qualitative analysis of interviews carried out with those who requested the rapid reviews and (2) a critical self-reflection by the authors (one researcher, three RSEs) on the applicability of EBP techniques to RSE work which weaves in insights from the broader EBP literature.


\section{Background}
\label{sec:background}

\textbf{The Future of Research Software Engineering.} In the past decade, the research software engineering movement has flourished as evidenced by the establishment of RSE groups at labs and universities\cite{katz2019research}\cite{milewicz2020department}\cite{cosden2023princeton}\cite{malviya2023research}, regional and national organizations for RSEs\cite{katz2018community}, and workshops and conferences for RSEs; there is growing recognition that RSEs play a vital role within the computational science and engineering workforce\cite{hacker2022building}\cite{bernholdt2022workshop}\cite{mundt2023public}. According to 2022 International RSE Survey, RSEs are employed at institutions across the world where they develop software, conduct and support research, and fill many other roles such as providing training and managing projects\cite{hettrick_simon_2022_7015772}. Now, over ten years out from when the term RSE was first coined, we must cast our gaze forward: how and in what directions must the field grow and mature?


Along these lines, a recent report by Lamprecht et al. has identified the future skill needs of RSEs as an open question for the community\cite{lamprecht2022we}. Already we see an evolving horizon that may impact RSEs' work, including an ever-shifting landscape of hardware and software\cite{yoshii2021does}\cite{van2020lessons}, the increasing importance of software security and sustainability\cite{meinel2022security}\cite{milewicz2022secure}\cite{morris2021understanding}, and the use of AI in scientific software development\cite{orlando2023assessing}\cite{piccolo2023many}. It is clear that the demand for software engineering expertise will continue to grow and change, and RSEs will need to keep pace. Many authors in recent years (including ourselves) have called for more training and mentorship to support RSEs' career development \cite{cohen2020four}\cite{cosden2023entry}\cite{milewicz2021mentorship}\cite{carver2022survey}\cite{trumbo2021learning}. Adding to this discussion, in our work we investigate evidence-based practice (EBP) as a pathway towards career-long learning, further professionalization of the field, and better practices in scientific software development. By incorporating a greater variety of high-quality evidence into their work, RSEs could make more informed and objective decisions about which approaches and techniques are most effective for a given task. Adoption of EBP could also help to promote a culture of continuous learning and improvement within the field, as practitioners actively seek out and evaluate new evidence in order to continually refine and improve their practices.


\textbf{Towards Evidence-Based RSE Practice.} The EBP paradigm represents both an ideological movement and a set of processes around the incorporation of research evidence into practice, beginning in the late 1980s and 1990s in medicine and subsequently spreading to other disciplines such as nursing, education, clinical psychology, and social work\cite{satterfield2009toward}. Around the same time, EBP began making inroads on software engineering starting with the work of Dyb\r{a}, Kitchenham, and J\o{}rgensen\cite{kitchenham2004evidence}\cite{dyba2005evidence}. As a \textit{way of thinking}, EBP promotes the idea that while nothing can take place of professionals' experiences and instincts, using evidence from research can help reduce bias and risk in decision-making and support continuous learning. Meanwhile, as a \textit{way of working}, EBP is centered around a five-step process for incorporating research evidence into practice, sometimes referred to as the five A's:

\begin{enumerate}
\item \textit{Ask}: Convert a relevant problem or information need into an answerable question.
\item \textit{Acquire}: Search the literature for the best available evidence to answer the question.
\item \textit{Appraise}: Critically appraise the evidence for its validity, impact, and applicability.
\item \textit{Apply}: Integrate the appraised evidence with practical experience and stakeholders' values and circumstances to make decisions about practice.
\item \textit{Analyze}: Finally, evaluate performance and seek ways to improve it.
\end{enumerate}

\begin{figure}
	\centering
	\includegraphics[width=0.8\linewidth]{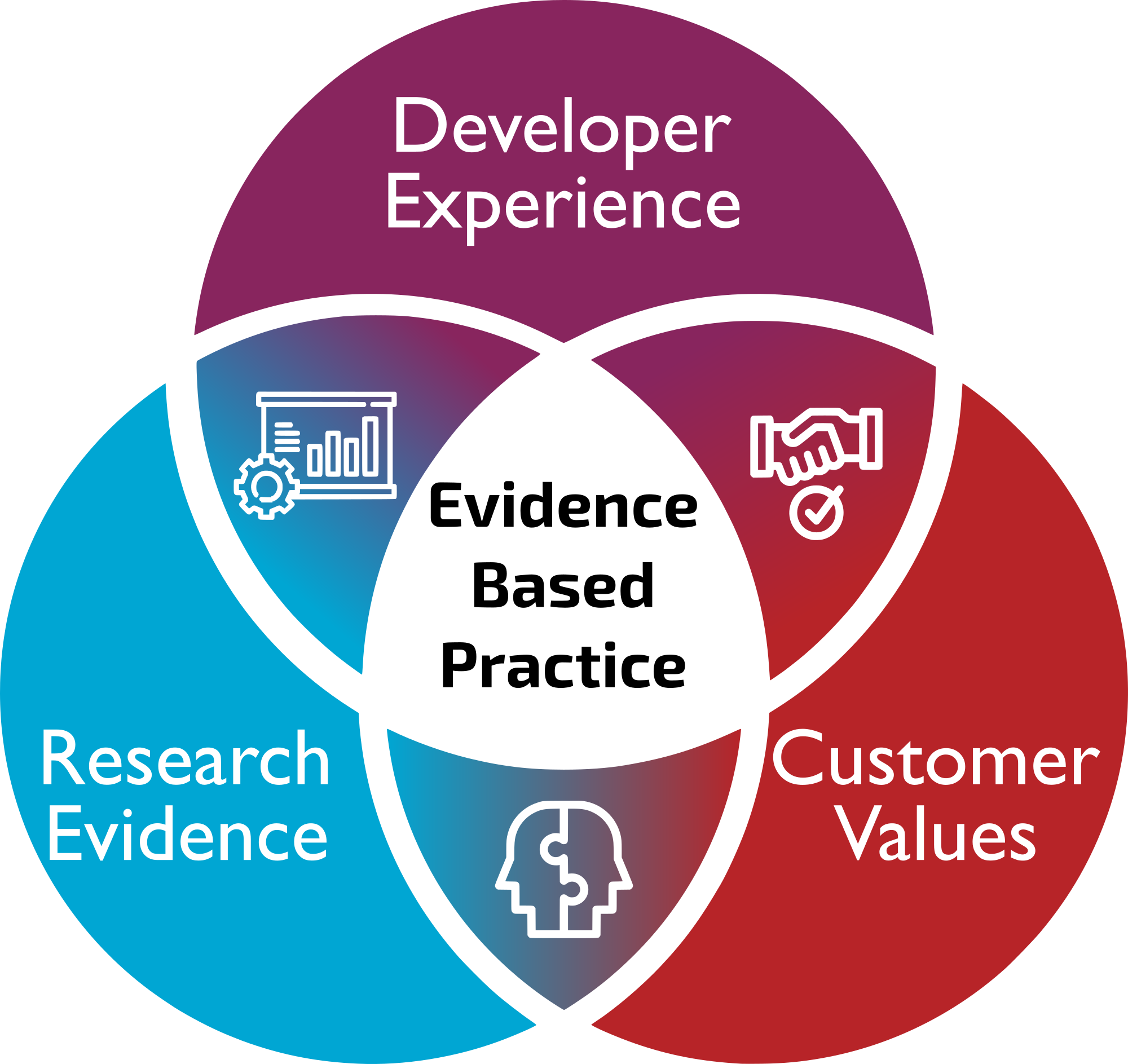}
	\caption{A diagram illustrating the key elements of evidence-based practice in software engineering. The goal of EBP is to unite practitioners' experiences and the needs and values of their customers with the depth and rigor of findings from research, enabling them to make better decisions and produce better software.}
	\label{fig:keyelements}
\end{figure}

Within this schema, EBP encompasses a variety of practices such as techniques for generating evidence of practical value (\textit{e.g.,} randomized controlled trials and meta-analyses), discovering evidence (\textit{e.g.,} literature review protocols), routines for using that evidence (\textit{e.g.,} practice guidelines and decision aids), and tools for auditing performance (\textit{e.g.,} stakeholder feedback and self-reflection tools). At present, certain EBP techniques have been widely adopted by software engineering (SE) researchers (especially literature reviews\cite{kitchenham2009systematic}), but less is known about their uptake among practitioners. While formally-trained software engineers learn about the latest research findings during their education, few are trained in EBP techniques to stay current on advances in their field after they graduate\cite{pizard2022longitudinal}. Indeed, a survey by Devanbu et al. found that developers frequently hold strong opinions based primarily on personal experience and what they hear from their peers and mentors, taking empirical research into consideration only a quarter of the time\cite{devanbu2016belief}. Le Goues et al. note that researchers are partly to blame because software engineering research papers---unlike works in, say, evidence-based medicine---may lack actionable advice and/or fail to convey the extent to which their findings can be trusted\cite{le2018bridging}. To be clear, none of this diminishes the accomplishments of the field. The work of software engineers is grounded in over five decades of excellence in software craftsmanship, and the widely-recognized best practices of the field are backed both by experience and research evidence. That being said, we see opportunities to further advance the field by enabling practitioners to engage with research more often in their daily operations. In short, while the concept of EBP has merit, more work is needed to adapt EBP practices and principles to the context of software engineering.

We argue that RSEs, perhaps moreso than other software engineering communities, are well-positioned to both benefit from and to champion the use of research evidence in software development practice. First, as Heroux has argued, software engineering and social science research could be a force multiplier for RSEs\cite{heroux2023research}; by applying the scientific method to the practice of scientific software development, we can improve upon the quality and rigor of our work and elevate respect for software craftsmanship. Here, we believe RSEs are especially capable: many come from backgrounds in math and science\cite{hettrick_simon_2022_7015772} and by definition work in interdisciplinary roles alongside other science and engineering professionals\cite{sims2022research}\cite{sufi2021rise}, meaning they are more likely than conventional software engineers to be conversant in the methods, language, and norms of scientific inquiry. Second, a key challenge for RSEs is communicating the importance of software engineering best practices to colleagues in other disciplines\cite{mundt2021working}. RSEs operate within cultures of science and engineering that value empiricism, and showing that their practice is grounded in over five decades of research could be a powerful tool for persuasion. Finally, RSEs adopting EBP techniques would help drive production of software engineering research that serves the needs of the RSE community\cite{milewicz2021building}; it would help software engineering researchers forge stronger ties to the RSE community in order to support and empower them.

\textbf{Rapid Reviews}. In this study we investigate the use of a particular EBP technique in an RSE context: rapid reviews. A \textit{rapid review protocol} is a systematic, time-boxed literature review designed to deliver evidence in a timely and accessible way\cite{khangura2012evidence}. Rapid reviews are motivated by practical problems and report results directly to practitioners in the field. To provide faster turnaround on results compared to full systematic reviews, rapid reviews omit and simplify certain steps such as by limiting the scope of the literature search, reducing the screening and quality appraisal steps, and presenting highlights from the literature without formal synthesis. Regarding the use of rapid reviews in software engineering, we owe much to the pioneering work in recent years by Cartaxo et al., who have explored the use of rapid reviews both in the software industry and in academia\cite{cartaxo2018role}\cite{cartaxo2019software}\cite{cartaxo2020rapid}.

Literature reviews in EBP encompass a spectrum of activities ranging from brief informal searches of online databases to months-long, full systematic reviews. Formal reviews tend to be more comprehensive, exhaustive, and provide higher quality results compared to informal searches, but are also more work-intensive and demand a substantial time commitment\cite{niazi2015systematic}. Rapid reviews are in the middle of that spectrum, providing faster turnaround compared to full reviews (\textit{e.g.,} within a two-week agile sprint) while offering a broader view of the evidence compared to informal searches. 

\section{Related Work}
\label{sec:relatedwork}

While EBP is common in other disciplines (e.g., healthcare~\cite{scurlock2014evidence}), there are few publications relating to the application of EBP in practice in SE contexts. Of note, Kasoju et al. in 2013 detailed an application of EBP (referred to as evidence-based software engineering, EBSE) in the automotive industry~\cite{kasoju2013analyzing}. They apply EBSE through a combination of case study, systematic literature review, and value stream analysis to identify strengths and challenges of software testing. As another example, Cartaxo et al. in 2018 conducted an Action Research protocol to help a software team improve customer communication issues through EBSE~\cite{cartaxo2018role}. Cartaxo and team were aiming to gauge the value of EBSE to this particular project and analyze the downstream effects of its usage, finding that the team started doing their own research after seeing how useful the process was in their previous problem. Finally, Pizard et al. in 2023 published the results of a field investigation of the effect of presenting the concept, values, and limitations of EBSE to members of a government agency on their long-term actions~\cite{pizard2023assessing}. The authors visited the agency officials 16 months after the initial presentation and found that, while the concept was interesting, the officials were not sure how to put it into practice given lack of support and resources but did take more care to attempt to search for evidence-based answers to problems in a less formalized manner. 

The work by Pizard et al. more extensively explores the barriers to adoption of EBP in SE than the other works; however, none of these address EBP from the context of research software engineering. This paper is novel in that we focus our lens on the application of EBP for RSEs, including both the perceived benefits and challenges.



\section{Methodology}


\subsection{Research Context}

Sandia National Laboratories is one of seventeen United States Department of Energy laboratories; the national lab system conducts research and development in areas such as energy, the climate and environment, national security, and health. At Sandia, the Department of Software Engineering and Research is a cross-functional group of software professionals which provides flexible, on-demand RSE staffing for development, consultation, and support to other departments within the Center for Computing Research\cite{milewicz2020department}\cite{willenbring2020department}. As mentioned in the introduction, the first author of this work is a software engineering researcher. Beginning in 2020, he began offering literature reviews as a service to colleagues within the department; staff are free to request a review to answer questions relevant to their practice.

\subsection{Rapid Review Approach}

\begin{figure*}[h]
	\centering
	\frame{\includegraphics[width=0.3\textwidth]{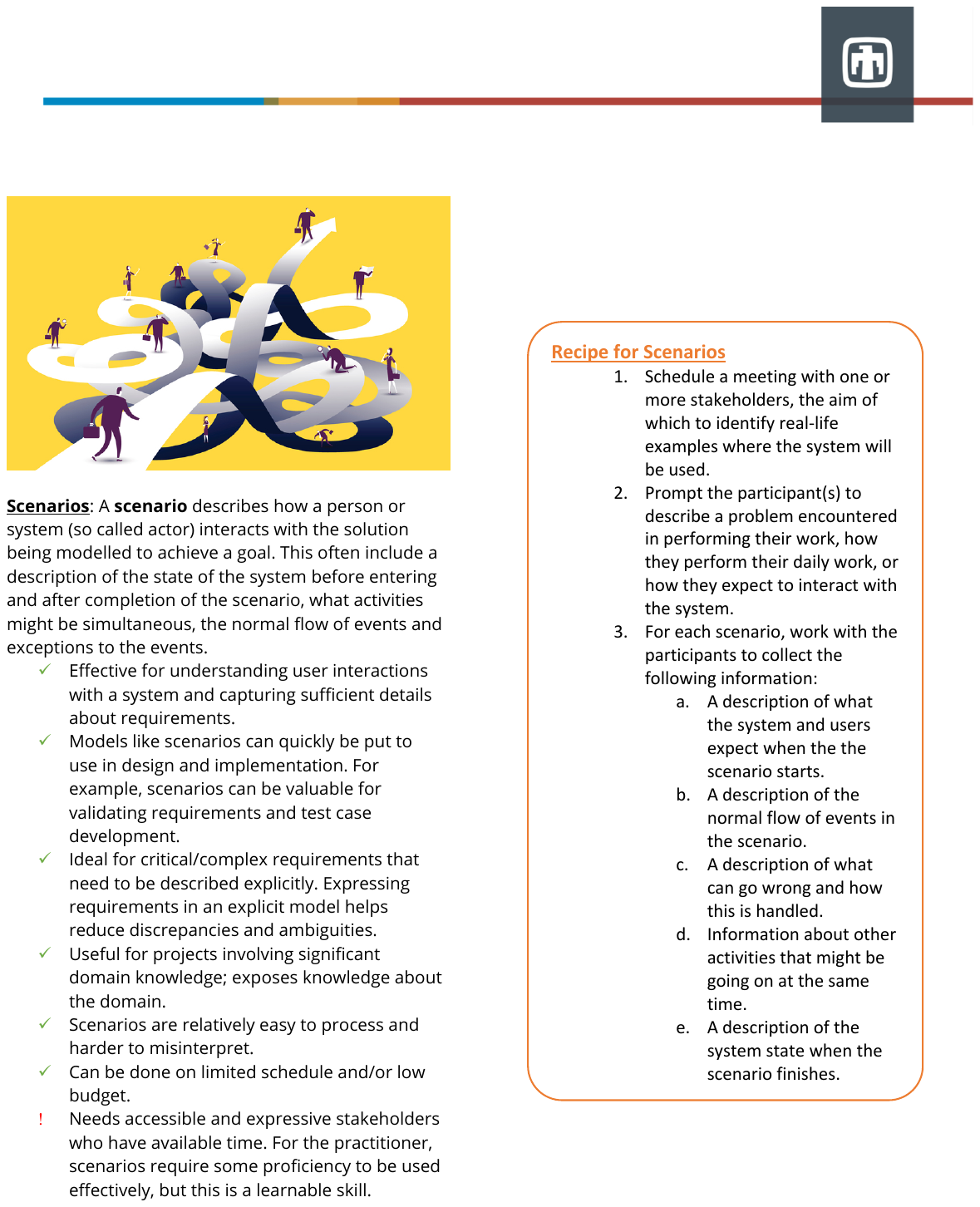}}
    \frame{\includegraphics[width=0.3\textwidth]{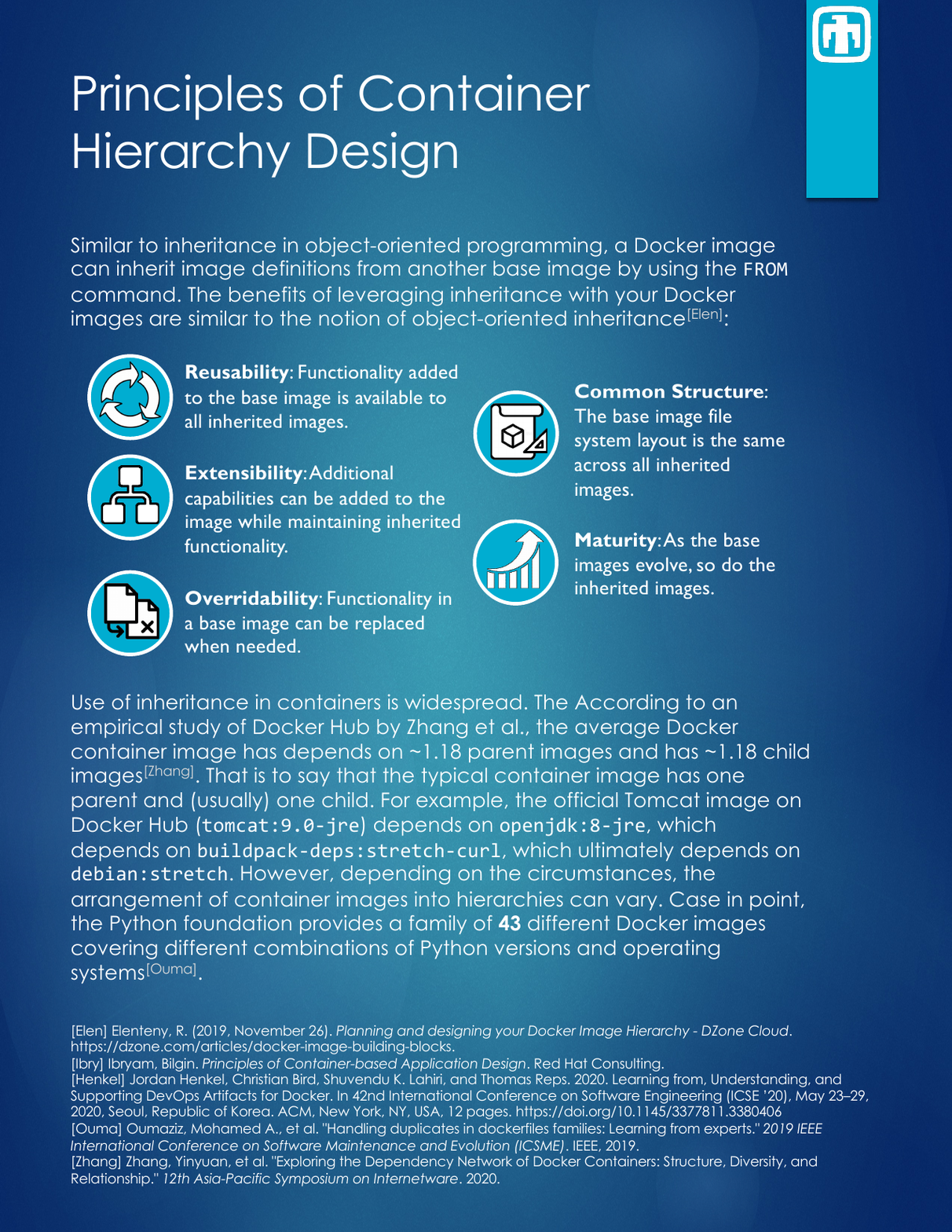}}
    \frame{\includegraphics[width=0.3\textwidth]{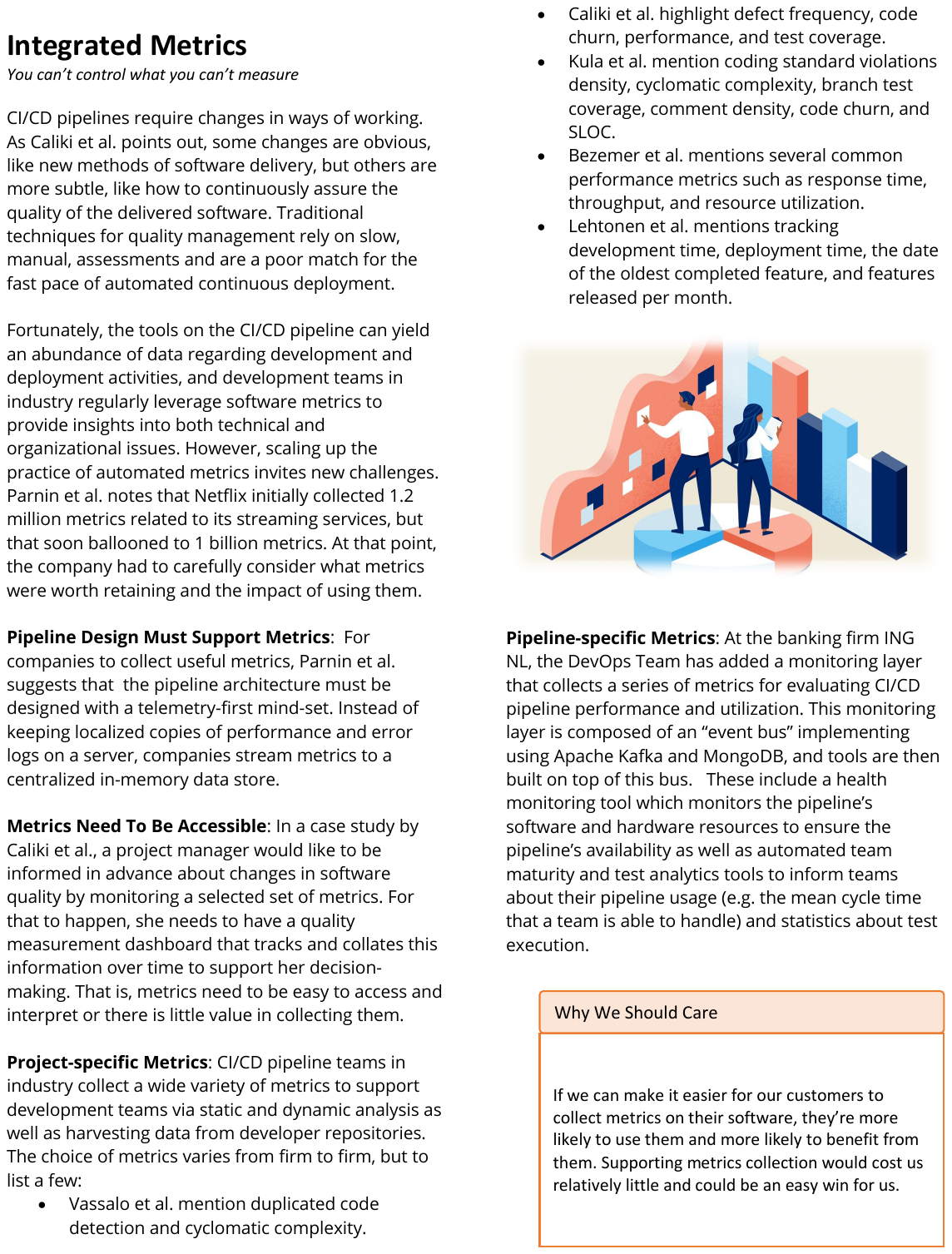}}
	\caption{From left to right, excerpts from evidence briefings resulting from rapid reviews \textit{RR2}, \textit{RR3}, and \textit{RR4}. The end goal of a rapid review is to translate findings from research literature into actionable guidance for practitioners.}
	\label{fig:examples}
\end{figure*}
\begin{table*}[h]
\centering
\caption{Summary of selected rapid reviews. For each review, we derived one or more answerable questions and mined the available literature using online academic search engines to find answers to those questions.}
\label{tab:rrsummary}
\begin{tabular}{@{}p{0.2\textwidth}p{0.35\textwidth}llp{0.1cm}@{}}
\toprule
Topic & Question(s) & Literature Selection Process & Feedback Interview\\ \midrule
\textit{RR1}: Software Quality Standards & Do different teams work better under different sets of quality standards? How do we ``right-size'' a software quality model? & (GScholar)1000\textgreater{}100\textgreater{}13\textgreater{}7 & Yes \\ \rowcolor[HTML]{EFEFEF} 
\textit{RR2}: Requirements Elicitation & What requirements techniques have evidence for their effectiveness, and when and where should they be applied, particularly in domain-specific and/or online/remote contexts? & (GScholar)1000\textgreater{}149\textgreater{}38\textgreater{}19 & Yes \\
\textit{RR3}: Software Containers & What are best practices in the design and maintenance of container image hierarchies? What are common use cases and requirements for containers in scientific computing? & (GScholar)1000\textgreater{}266\textgreater{}53\textgreater{}33 & No \\ \rowcolor[HTML]{EFEFEF} 
\textit{RR4}: CI/CD Pipelines & What successes have teams in industry realized in the area of CI/CD pipelines, in particular industries that have complex workflows and heterogeneous architectures? & (GScholar)1000\textgreater{}315\textgreater{}66\textgreater{}39 & Yes \\
\textit{RR5}: Web Crawling & What is the current state of the art in web crawling technologies, particularly in the areas of intelligent and deep web crawling? What workshops, conferences, and journals feature web crawling research? How active is this research community, and what topics do they publish on? & (ACM+IEEE)284\textgreater{}183 & Yes \\ \rowcolor[HTML]{EFEFEF} 
\textit{RR6}: Software Quality Incentivization & What strategies are recommended to incentivize software teams to invest in software quality? What are their benefits and drawbacks? & (GScholar)1000\textgreater{}42\textgreater{}26 & No \\ \bottomrule
\end{tabular}
\end{table*}

The rapid review approach our team uses was inspired by the work of Cartaxo et al.\cite{cartaxo2018role}, and consists of the following steps:

\begin{enumerate}
    \item (Ask) We hold a meeting (typically 15-30 minutes long) to understand participants' information needs. These needs are distilled down into an answerable question which can then be converted into a search query, and the query is agreed upon by all involved. 
    \item (Acquire) Using one or more academic search engines, the researcher executes the search query and collects a set of search results. In the case of the rapid reviews described in this paper, we used Harzing's Publish or Perish (Version 7.22) to mine the top 1000 results from Google Scholar; the exception to this was RR5 where the ACM Digital Library and IEEE Xplore were used in lieu of Google Scholar. This results in a spreadsheet of papers which then need to be filtered.
    \item (Appraise) Papers in the search results are then downselected in iterative phases according to availability (papers have to be in English, peer-reviewed, and accessible via our institution's subscriptions), then relevance (whether the abstract pertains to the answerable question), then quality (papers were assigned scores from 1-5 based on the perceived usefulness of the content).
    \item (Apply) The researcher takes the top-ranked papers and analyzes them to extract useful content; we use open coding techniques (see \cite{holton2007coding}) to label and extract themes from the literature. These results are then distilled into an evidence briefing, a narrative summary with useful guidelines and information for practitioners. The researcher then holds a meeting to summarize the findings in the briefing with the participants. 
    \item (Analyze) In most cases, around 1-2 months after the briefing, we hold a retrospective meeting to assess the usefulness and impact of the review to get feedback on how the review process could be improved in the future.
\end{enumerate}

In this paper, we present an analysis of six selected rapid reviews our team has carried out; a summary of these reviews is provided in Table \ref{tab:rrsummary}. For all six reviews we kept meticulous notes on the review process, and for four of the six reviews we performed a retrospective interview several months after the delivery of the evidence briefing.  As this is an experience report and all involved are members of the same team, it was determined that ethics board approval was not required for this research. In the interest of privacy, however, we use pseudonyms when quoting participants' feedback. While we do not make interview data available, copies of the evidence briefings and sample data illustrating the rapid review process are available on Zenodo\footnote{\url{https://doi.org/10.5281/zenodo.8169610}}\cite{EBPDataset}. For \textbf{RQ1}, we used open coding techniques on the interview data to draw out themes and computed estimates on the time and effort required to perform the literature reviews. For \textbf{RQ2}, we reflected on the challenges that RSEs may face in adopting EBP techniques, and then performed an informal search of the broader EBP literature to identify strategies that have been used in other disciplines to help overcome those challenges.

\section{Results}
\label{sec:results}

\subsection{RQ1: What are the strengths and limitations of rapid reviews in RSE contexts?}
\subsubsection{When should rapid reviews be performed?}

As noted by Freund, software engineers are knowledge workers who frequently seek out new information to help make decisions and solve problems\cite{freund2015contextualizing}. Information can come from many different sources, such as hands-on learning, reaching out to colleagues, delving into code and documentation, and consulting technical literature including blog posts, books, and peer-reviewed articles. On our team, participants commented on how the reviews helped them to get the ``lay of the land'' for a topic and/or to help free them from indecision or ``analysis paralysis'':

\begin{quote}
    \faCommentingO \hspace{1pt} \textbf{Henry (RR5)}: Yes, I found it very helpful. It is a useful technique, a useful process for very quickly getting a sense of what's out there. I found it very useful because you drew attention to papers and topic areas that I wasn't necessarily aware of.
\end{quote}
\begin{quote}
    \faCommentingO \hspace{1pt} \textbf{Cate (RR1)}: One of the things that I really liked about it is that you gave a broad scope at the start. Here are these categories that may have some influence. I didn't end up using all of them because they weren't all relevant. But it was good to widen my own understanding of these topics. 
\end{quote}
\begin{quote}
	\faCommentingO \hspace{1pt} \textbf{Anthony (RR4)}: [Rapid reviews] help us get unstuck. We could have these conversations that drag on forever. Should we look into something or not? Then 48 hours later, [our staff researcher] is in our inbox with an informative analysis of the topic. That kept us from heming and hawing about it. That got us off the snide, and we were able to make progress.
\end{quote}

In particular, all six rapid reviews involved mapping out the solution space for a given problem. For four of the six reviews (\textit{RR2}, \textit{RR3}, \textit{RR4}, and \textit{RR6}), the goal was to compare and contrast different software engineering approaches for a task. In some cases, participants wanted to learn about what underlying factors influenced use of software engineering tools and practices, like what use cases scientific software developers have for containers (\textit{RR3}) or what dimensions of software team characteristics inform the practices they do or don't follow (\textit{RR1}). Meanwhile, in three cases participants wanted to know what the academic consensus was on a topic: what topics researchers are pursuing in algorithm development (\textit{RR5}), what software quality incentivization strategies are backed by empirical data (\textit{RR6}), and what evidence there is for the effectiveness of different requirements gathering techniques (\textit{RR2}).

Another common theme was that scientific software development is relatively understudied compared to conventional software, and as such a rapid review for RSE work may require an assessment of the transferability of the evidence to scientific software development. Several participants reflected on this aspect of the rapid review process:

\begin{quote}
    \faCommentingO \hspace{1pt} \textbf{Jack (RR4)}: I want to be able to lean on the results of this rapid review to tell me what do I need to be doing with continuous integration and deployment. For a given project, what should I prioritize? How do I get the best return on my investment, the greatest amount of improvement for the least amount spent? [...] At the same time, the results of this literature search need to be applicable to our needs, in our community. We have long build and test times, heterogeneous and specialized hardware, and very complex workflows.
\end{quote}
\begin{quote}
    \faCommentingO \hspace{1pt} \textbf{Cate (RR1)}: How do we 'right-size' a software quality model? How do we do a software quality standardization effort that is actually right-sized for the culture, teams, and what-not that we have here? [...] I need to figure out how to tailor [these findings] to my center, my population of 200 people, based on, say, this study of 175 information companies.
\end{quote}

That is, what works for industry may or may not be suitable for a scientific software project. Because rapid reviews are designed not just to collect research evidence but to interpret it and make it useful, the approach appears to be well-suited for recontextualizing industry practices in a systematic way. In producing the evidence briefing, the author must weigh the recommendations from the literature against their professional experience in the RSE domain.

\MyBox{\textbf{RQ1.1}: Our team found rapid reviews to be particularly helpful in \textbf{mapping out the solution space} for specific software engineering problems and \textbf{assessing the transferability of solutions} in the literature to scientific software development contexts.}

\subsubsection{How long do rapid reviews take to perform?}

We did not collect precise data on the time required to complete each rapid review, though each rapid review was carried out over the course of 3-5 business days. As a ``back of the envelope'' estimate, it took the first author 10-15 seconds to review each title and abstract for relevance, 1-2 minutes to assess the overall quality and applicability of an article based on the text, and then 10-20 minutes to read through and label the content of that article. Based on the rapid reviews presented in this paper where Google Scholar was used, it took 2-4 hours to mark 1000 articles according to relevance, 4-5 hours to downselect an average of 223 potentially relevant articles, and finally 4-8 hours to analyze an average of 24 high-quality sources. Writing each evidence briefing then took 2-3 hours. 

In total, each rapid review required 12-20 hours of labor by a single trained researcher with experience in reading and interpreting research literature and spread out over 3-5 days. This timeline tracks with rapid reviews in medicine, where Hartling et al. report that the shortest of rapid reviews are completed in a week or less (though 5-8 weeks is typical for published rapid reviews)\cite{hartling2015taxonomy}; this is significantly shorter than a full systematic review, which Borah et al. estimate can take 26-67 weeks using traditional methods\cite{borah2017analysis}.

\MyBox{\textbf{RQ1.2}: Each rapid review reported in this paper, as performed by a trained software engineering researcher, required \textbf{12-20 hours of labor} over the course of three to five days to complete. This is much shorter than a full systematic review (which can take 26-67 weeks to perform). }

\subsubsection{What are the limitations of rapid reviews?}

First and foremost, the time estimates given above assume that the person performing the search (1) has experience in performing literature reviews and (2) is familiar with software engineering literature in particular. Our RSE team benefits from having staff researchers on-hand who are able to fill this role, but this may be uncommon elsewhere. The question then is whether RSEs can perform these rapid reviews themselves. As we noted in Section \ref{sec:background}, many RSEs come from backgrounds in science and engineering rather than specifically software development. This suggests that while they may know how to perform literature reviews, they are less equipped to navigate academic software engineering literature. As Cate explains,

\begin{quote}
    \faCommentingO \hspace{1pt} \textbf{Cate (RR1)}: Speaking as someone who didn't come from a software engineering background, if I were trying to sift through all of this knowledge on my own, I wouldn't even know what I was looking for. It's extremely helpful to me to know that I have someone that I can reach out to, someone who knows these things, how to look, where to look, and who can evaluate which resources are reliable or unreliable, and who can come back not with just one suggestion but a really thoroughly wide spread net of ``I researched this topic, identified these resources, and this is the consensus based on the literature I have found for you.''
\end{quote}

Conversely, RSEs who come from conventional software engineering backgrounds may lack training in conducting literature reviews. Along these lines one participant, Drew, suggested that we document the process of how to find and interpret academic literature for the benefit of junior staff on the project he leads, some of whom were recently hired in from industry:

\begin{quote}
	\faCommentingO \hspace{1pt} \textbf{Drew (RR5)}: Yeah, I'd love for you to have a place to document essentially the resources one has or the processes one goes through to go through or acquire academic research. Someone at my level should have more of those skills by now, at my level. But there are people much more junior who have never been asked to do that kind of stuff yet. A wiki page that talks about what you can provide, rapid reviews, your process, and the other part about how does a team member do their own research.
\end{quote}

Meanwhile, another theme we found was that there are often assumptions encoded into software engineering research literature---which primarily considers the needs of conventional industry---that may not hold for research software engineering:

\begin{quote}
    \faCommentingO \hspace{1pt} \textbf{Allen (RR4)}: If we wanted to explore containerization further, we would have to build a business case for it, and that would include finding or producing evidence that containerized execution can work for scientific software applications more generally and then for our customers’ applications specifically. There’s no sense in rolling out a new technology until we’re confident that it’ll work. 
\end{quote}
\begin{quote}
	\faCommentingO \hspace{1pt}  \textbf{Sebastian (RR4)}: I appreciated [the sources] calling out that you need a dedicated pipeline team. It’s rare to find a code team with dedicated DevOps staff, and even then, it’s rare to have more than one person dedicated to DevOps. But if you want to mature in this direction, you need dedicated staff.
\end{quote}

In Allen's case, the fact there are relatively few studies on the performance of containers in scientific computing contexts compared with conventional industry contexts makes it difficult to draw confident conclusions from the literature; for example, a study on the performance of containerized microservices on a distributed cloud tells us little about the use of containers for an MPI-based HPC application\footnote{It is worth noting, however, that this observation inspired us to do a more in-depth rapid review focused on containers and their potential for use in scientific computing (\textit{RR3}) That review yielded more sources concerning the use of containers in HPC which were not present in our review of CI/CD pipelines (\textit{RR4}).}. Meanwhile, for Sebastian, the literature on CI/CD pipelines recommends having a dedicated DevOps team to maintain a pipeline, but scientific software projects are lucky to have even one such professional on their project---let alone an entire team.

\MyBox{\textbf{RQ1.3}: In terms of limitations, performing rapid reviews requires both \textbf{the skills to conduct literature reviews} and \textbf{familiarity with the research literature} in software engineering. Moreover, rapid reviews, like any EBP technique applied to RSE work, are limited by the fact that \textbf{scientific software development is understudied}.}

\subsection{RQ2: What are the challenges for RSEs to adopt EBP techniques more generally? What strategies would help address those challenges?}

While the focus of this article so far has been on rapid reviews, our interests in EBP are much broader, and our position is that the evidence-based paradigm is generally useful and that the RSE community should employ EBP techniques where appropriate. We recognize, however, that this will not happen overnight, and based on our experiences we anticipate challenges to EBP adoption. To that end, we offer a critical self-reflection on those experiences, and we present insights from other fields where EBP has been successfully adopted to suggest strategies to address potential roadblocks.

\subsubsection{Challenge: RSEs Need Training}
While our team was able to complete each rapid review in a timely manner, this was due to the fact that the effort was led by a trained researcher who was already familiar with the breadth of the software engineering literature. As we mentioned previously, an unprepared RSE would likely struggle to do the same. Interpreting research literature requires competencies that not all RSEs have. As Cosden et al. has pointed out, a growing number of new RSEs are coming from industry rather than domain science or math backgrounds\cite{cosden2023entry}. 

\textbf{Insights}: A review by Rousseau and Gunia finds that practitioners are most likely to implement EBP when they have ``the ability (foundational and functional competencies), motivation (behavioral beliefs, perceived behavioral control, and normative beliefs), and opportunity (support that overcomes barriers) to do so''\cite{rousseau2016evidence}. As mentioned in Section \ref{sec:relatedwork}, findings by Pizard et al. suggest that this ability can be taught to software engineers\cite{pizard2022longitudinal}. Regarding motivation, potential solutions include having an EBP mentor\cite{melnyk2004nurses} and offering training that focuses on building up confidence and self-efficacy in EBP techniques\cite{kiss2010self}. Additionally, Leeman et al. argues that motivation can be strengthened through ``capacity building'' strategies such as establishing peer support networks and packaging evidence-based practices into standardized formats that they can easily adapt and use in their daily work\cite{leeman2017developing}. Finally, the opportunity to apply has been linked to having on-the-job autonomy and flexibility to experiment with new approaches\cite{belden2012effect}, having the authority to act on evidence\cite{dalheim2012factors}, and having support from supervisors\cite{melnyk2012state}.

\MyBox{\textbf{RQ2.1}: While there is emerging evidence that \textbf{EBP techniques can be taught to software engineers}, studies suggest that the \textbf{motivation and opportunity} to apply EBP are equally important. Approaches for increasing motivation include encouraging mentorship, offering training to increase confidence and self-efficacy in EBP, establishing support networks, and packaging evidence-based resources to make it as easy as possible for RSEs to engage in EBP. Meanwhile, approaches for creating opportunities include securing buy-in from managers/supervisors, the authority to act on evidence, and the autonomy and flexibility to experiment.
}

\subsubsection{Challenge: Research is Incomplete}

In making the case for EBP, there is an implicit assumption that research evidence exists that would support RSEs' specific needs. This was true for the particular rapid reviews presented in this paper, but it is guaranteed that there are many more worthwhile questions RSEs have that haven't received scholarly attention. In general, research literature tends to have a bias towards new and innovative topics while countless practical questions have never been rigorously studied. This is doubly true for studies concerning scientific software development, given that most software engineering research focuses primarily on the needs of the software industry\cite{milewicz2021building}. Where studies do exist for a given question, they rarely provide clear-cut guidance; readers have to carefully parse the caveats and limitations of research studies when drawing conclusions\cite{wohlin2021challenges}. Moreover, as we found in our exploration of rapid reviews, there is a need to filter and contextualize findings to suit the particulars of RSE practice.

\textbf{Insights}: This problem is not unique to software engineering. As one physician put it, ``We don’t have solid evidence for the majority of the care we provide, and no concrete plan for remedying that problem exists. Such a solution would take decades to achieve''\cite{callaham2015expert}. This is not, however, a condemnation of EBP: the foundation of EBP is professional experience and intuition, and research can complement but never replace those things\cite{klein2016can}. That being said, RSEs would benefit from having more research studies tailored to their needs. Closing the research gap will require a closer working relationship between software engineering researchers and RSEs. Goldstein et al. have found that establishing researcher-practitioner networks can help promote this kind of engagement and collaboration\cite{goldstein2018practice}. Practice-Based Research Networks (PBRNs) in medicine are multi-institutional partnerships which (1) give researchers access to sites to conduct studies and (2) give practitioners access to leading experts who can help support improvement activities. Similarly, professional societies can act as liaisons and facilitators for the development of evidence-based recommendations and practice\cite{neilson2015moving}. 

\MyBox{\textbf{RQ2.2}: Given the complex and evolving nature of RSE work, research evidence is virtually guaranteed to be limited and incomplete; these circumstances have not stopped the adoption of EBP in other fields. There is, however, a clear need for more research focused on RSEs' needs. This will require \textbf{a closer working relationship between software engineering researchers and RSEs}, possibly via organizational models like Practice-Based Research Networks and facilitated networking by national RSE organizations.}


\subsubsection{Challenge: Convincing Others to Adopt Best Practices}

Practice based on evidence is often portrayed as an alternative to practice based solely on authority or tradition\cite{rodwin2001commentary}\cite{gambrill2018evidence}. In other words, the evidence-based paradigm harbors an expression of hope that decision-making guided by empirical data will cut through politics, personal biases, and differences of opinion. In truth, we recognize RSEs are often embedded in multi-disciplinary science and engineering teams who may not share the same outlook or priorities with regards to software engineering best practices\cite{mundt2021working}. While RSEs can build an evidence-based case for a course of action, they themselves are usually not the ones that need convincing. For example, efforts to standardize and incentivize software quality among scientific software projects (\textit{RR2} and \textit{RR6}) or to set up containerized deployments and CI/CD pipelines for them (\textit{RR3} and \textit{RR4}) may all require winning over potentially reluctant teams.

\textbf{Insights}: Comparable situations are frequent in healthcare settings; for example, each member of a multi-disciplinary care team can have a different opinion on how to treat a patient based on their distinct education, training, and experience, and this can lead to conflict\cite{jacobs2015principles}. Studies suggest that may stem in part from that a lack of shared clarity around roles and expertise\cite{kutash2014quality}. A review by Mathieson et al. finds that this can can pose a barrier to EBP adoption, noting that ``in order to implement an innovation within a multi-disciplinary team, adopters need to understand each other’s roles and workload''\cite{mathieson2019strategies}. This underscores the need for dialogue on multi-disciplinary teams to build this shared understanding of the knowledge and skills and to improve the working relations.

\MyBox{\textbf{RQ2.3}: RSEs often work alongside science and engineering professionals who have different values and priorities regarding software engineering best practices. To fully realize the value of EBP on their teams, studies recommend \textbf{promoting dialogue to establish a shared clarity and a positive attitude towards the knowledge and skills that each person brings to the team.}}

\section{Discussion}

In rallying for EBP techniques, we are not suggesting that current software engineering practice is ungrounded or not rigorous. Rather, we see evidence-based practice as a pathway towards novel, useful ways of working in software engineering, and we believe the research software engineering community could play a key role in proving it out. The research software engineering movement has been successful in narrowing the historical ``chasm'' between software engineers and domain researchers\cite{kelly2007software}, bringing a heritage of excellence to scientific computing. Moving forward, we see opportunities to enable RSEs to engage with research more often in their daily operations. Evidence-based software engineering approaches, we argue, could help promote career-long learning, further the professionalization of the field, and encourage the adoption of software engineering best practices in scientific software development.

\section{Threats to Validity}

In interpreting and generalizing the findings of this research, it is important to consider several potential threats to validity. First and foremost, this study represents an experience report on the use of rapid reviews rather than a controlled experiment. Experience reports provide rich, context-specific insights, but they also carry inherent biases: what worked for our particular RSE team may or may not work for another team. Second, this study provides guidance based on a synthesis of the existing EBP literature outside software engineering; future research is needed to confirm the transferability of those findings to RSE contexts.

\section{Conclusion}

Our study represents a first-in-kind effort to apply the evidence-based paradigm to research software engineering. First, we investigated the use of rapid reviews as an EBP technique to translate research evidence to RSE practice. We found rapid reviews to be an efficient method for mapping out the solution space for different software engineering problems and to assess the transferability of findings from the literature to RSE contexts. We noted several limitations, however: the need for training to perform literature reviews, the incompleteness of research, and the need for buy-in from non-RSEs to implement evidence-based best practices on teams. Based on our experiences and a review of the broader EBP literature, we derived a set of recommendations for how to further EBP adoption among RSEs. This includes (1) creating support networks, providing training, and securing buy-in from supervisors to pursue EBP, (2) establishing a closer working relationship between software engineers and RSEs to close research gaps, and (3) building shared clarity between RSEs and the communities they serve on the skills and knowledge that RSEs bring to the table through dialogue. In future work, we hope to collect data on information needs of RSEs to guide software engineering researchers and to develop training materials on evidence-based methods geared towards RSEs.

\bibliographystyle{IEEEtran}
\bibliography{evidenceBasedPractice}

\begin{thebibliography}{10}
\providecommand{\url}[1]{#1}
\csname url@samestyle\endcsname
\providecommand{\newblock}{\relax}
\providecommand{\bibinfo}[2]{#2}
\providecommand{\BIBentrySTDinterwordspacing}{\spaceskip=0pt\relax}
\providecommand{\BIBentryALTinterwordstretchfactor}{4}
\providecommand{\BIBentryALTinterwordspacing}{\spaceskip=\fontdimen2\font plus
\BIBentryALTinterwordstretchfactor\fontdimen3\font minus
  \fontdimen4\font\relax}
\providecommand{\BIBforeignlanguage}[2]{{%
\expandafter\ifx\csname l@#1\endcsname\relax
\typeout{** WARNING: IEEEtran.bst: No hyphenation pattern has been}%
\typeout{** loaded for the language `#1'. Using the pattern for}%
\typeout{** the default language instead.}%
\else
\language=\csname l@#1\endcsname
\fi
#2}}
\providecommand{\BIBdecl}{\relax}
\BIBdecl

\bibitem{baxter2012research}
R.~Baxter, N.~C. Hong, D.~Gorissen, J.~Hetherington, and I.~Todorov, ``The
  research software engineer,'' in \emph{Digital Research Conference, Oxford},
  2012, pp. 1--3.

\bibitem{dyba2005evidence}
T.~Dyba, B.~A. Kitchenham, and M.~Jorgensen, ``Evidence-based software
  engineering for practitioners,'' \emph{IEEE software}, vol.~22, no.~1, pp.
  58--65, 2005.

\bibitem{milewicz2020department}
R.~Milewicz, J.~Willenbring, and D.~Vigil, ``Research, develop, deploy:
  Building a full spectrum software engineering and research department,''
  2020.

\bibitem{willenbring2020department}
\BIBentryALTinterwordspacing
J.~M. Willenbring and R.~Milewicz, ``Moving forward together: How a software
  engineering department can impact developer productivity in a research
  organization.'' \emph{The 2020 Collegeville Workshop on Scientific Software},
  6 2020. [Online]. Available: \url{https://www.osti.gov/biblio/1806263}
\BIBentrySTDinterwordspacing

\bibitem{cartaxo2018role}
B.~Cartaxo, G.~Pinto, and S.~Soares, ``The role of rapid reviews in supporting
  decision-making in software engineering practice,'' in \emph{Proceedings of
  the 22nd International Conference on Evaluation and Assessment in Software
  Engineering 2018}, 2018, pp. 24--34.

\bibitem{katz2019research}
D.~S. Katz, K.~McHenry, C.~Reinking, and R.~Haines, ``Research software
  development \& management in universities: case studies from manchester's
  rsds group, illinois' ncsa, and notre dame's crc,'' in \emph{2019 IEEE/ACM
  14th International Workshop on Software Engineering for Science
  (SE4Science)}.\hskip 1em plus 0.5em minus 0.4em\relax IEEE, 2019, pp. 17--24.

\bibitem{cosden2023princeton}
I.~A. Cosden, ``The princeton university rse group model: Operational and
  organizational approaches,'' \emph{Computing in Science \& Engineering},
  2023.

\bibitem{malviya2023research}
A.~Malviya-Thakur, D.~E. Bernholdt, W.~F. Godoy, G.~R. Watson, M.~Doucet, M.~A.
  Coletti, D.~M. Rogers, M.~McDonnell, J.~J. Billings, and B.~Maccabe,
  ``Research software engineering at oak ridge national laboratory,''
  \emph{Computing in Science \& Engineering}, 2023.

\bibitem{katz2018community}
D.~S. Katz, L.~C. McInnes, D.~E. Bernholdt, A.~C. Mayes, N.~P.~C. Hong,
  J.~Duckles, S.~Gesing, M.~A. Heroux, S.~Hettrick, R.~C. Jimenez
  \emph{et~al.}, ``Community organizations: Changing the culture in which
  research software is developed and sustained,'' \emph{Computing in Science \&
  Engineering}, vol.~21, no.~2, pp. 8--24, 2018.

\bibitem{hacker2022building}
T.~Hacker, P.~Smith, D.~Brunson, L.~Arafune, T.~Cheatham, and E.~Deelman,
  ``Building the research innovation workforce: Challenges and recommendations
  from a virtual workshop to advance the research computing community,'' in
  \emph{Practice and Experience in Advanced Research Computing}, 2022, pp.
  1--7.

\bibitem{bernholdt2022workshop}
\BIBentryALTinterwordspacing
D.~E. Bernholdt, J.~Cary, M.~Heroux, and L.~C. McInnes, ``The science of
  scientific software development and use,'' 5 2022. [Online]. Available:
  \url{https://www.osti.gov/biblio/1846008}
\BIBentrySTDinterwordspacing

\bibitem{mundt2023public}
M.~Mundt, K.~Beattie, J.~Bisila, C.~Ferenbaugh, W.~Godoy, R.~Gupta, J.~Guyer,
  M.~Kiran, A.~Malviya-Thakur, R.~Milewicz \emph{et~al.}, ``For the public
  good: Connecting, retaining, and recognizing current and future rses at
  national organizations,'' \emph{Computing in Science \& Engineering}, 2023.

\bibitem{hettrick_simon_2022_7015772}
\BIBentryALTinterwordspacing
S.~Hettrick, R.~Bast, S.~Crouch, C.~Wyatt, O.~Philippe, A.~Botzki, J.~Carver,
  I.~Cosden, F.~D'Andrea, A.~Dasgupta, W.~Godoy, A.~Gonzalez-Beltran,
  U.~Hamster, S.~Henwood, P.~Holmvall, S.~Janosch, T.~Lestang, N.~May,
  J.~Philips, N.~Poonawala-Lohani, P.~Richmond, M.~Sinha, F.~Thiery,
  B.~Werkhoven, and Q.~Zhang, ``International rse survey 2022,'' Aug. 2022, {If
  you use this dataset, please cite it using the metadata from this file.}
  [Online]. Available: \url{https://doi.org/10.5281/zenodo.7015772}
\BIBentrySTDinterwordspacing

\bibitem{lamprecht2022we}
A.-L. Lamprecht, C.~Martinez-Ortiz, M.~Barker, S.~L. Bartholomew, J.~Barton,
  N.~C. Hong, J.~Cohen, S.~Druskat, J.~Forest, J.-N. Grad \emph{et~al.}, ``What
  do we (not) know about research software engineering?'' \emph{Journal of Open
  Research Software}, vol.~10, 2022.

\bibitem{yoshii2021does}
K.~Yoshii, ``What does the post-moore era mean for research software
  engineering?'' \emph{Research Software Engineers in HPC (RSE-HPC-2021)},
  2021.

\bibitem{van2020lessons}
B.~van Werkhoven, W.~J. Palenstijn, and A.~Sclocco, ``Lessons learned in a
  decade of research software engineering gpu applications,'' in
  \emph{Computational Science--ICCS 2020: 20th International Conference,
  Amsterdam, The Netherlands, June 3--5, 2020, Proceedings, Part VII 20}.\hskip
  1em plus 0.5em minus 0.4em\relax Springer, 2020, pp. 399--412.

\bibitem{meinel2022security}
\BIBentryALTinterwordspacing
M.~Meinel and M.~Stoffers, ``Revisiting secure software engineering for
  research software,'' in \emph{RSE Conference 2022}, September 2022. [Online].
  Available: \url{https://elib.dlr.de/188479/}
\BIBentrySTDinterwordspacing

\bibitem{milewicz2022secure}
R.~Milewicz, J.~Carver, S.~Grayson, and T.~Atkison, ``A secure future for
  open-source computational science and engineering,'' \emph{Computing in
  Science \& Engineering}, vol.~24, no.~4, pp. 65--69, 2022.

\bibitem{morris2021understanding}
L.~Morris, ``Understanding software sustainability in the field of research
  software engineering,'' Ph.D. dissertation, University of Huddersfield, 2021.

\bibitem{orlando2023assessing}
G.~Orlando, ``Assessing chatgpt for coding finite element methods,'' no.
  MOX-Report No. 31/2023, 2023.

\bibitem{piccolo2023many}
S.~R. Piccolo, P.~Denny, A.~Luxton-Reilly, S.~Payne, and P.~G. Ridge, ``Many
  bioinformatics programming tasks can be automated with chatgpt,'' \emph{arXiv
  preprint arXiv:2303.13528}, 2023.

\bibitem{cohen2020four}
J.~Cohen, D.~S. Katz, M.~Barker, N.~C. Hong, R.~Haines, and C.~Jay, ``The four
  pillars of research software engineering,'' \emph{IEEE Software}, vol.~38,
  no.~1, pp. 97--105, 2020.

\bibitem{cosden2023entry}
I.~A. Cosden, K.~McHenry, and D.~S. Katz, ``Research software engineers: Career
  entry points and training gaps,'' \emph{Computing in Science \& Engineering},
  2023.

\bibitem{milewicz2021mentorship}
R.~Milewicz and M.~Mundt, ``An exploration of the mentorship needs of research
  software engineers,'' \emph{Research Software Engineers in HPC
  (RSE-HPC-2021)}, 2021.

\bibitem{carver2022survey}
J.~C. Carver, N.~Weber, K.~Ram, S.~Gesing, and D.~S. Katz, ``A survey of the
  state of the practice for research software in the united states,''
  \emph{PeerJ Computer Science}, vol.~8, p. e963, 2022.

\bibitem{trumbo2021learning}
D.~Trumbo and R.~Milewicz, ``Poster: Towards a culture of continuous learning
  and improvement within rse teams.''\hskip 1em plus 0.5em minus 0.4em\relax
  2021 Collegeville Workshop on Scientific Software, 2021.

\bibitem{satterfield2009toward}
J.~M. Satterfield, B.~Spring, R.~C. Brownson, E.~J. Mullen, R.~P. Newhouse,
  B.~B. Walker, and E.~P. Whitlock, ``Toward a transdisciplinary model of
  evidence-based practice,'' \emph{The Milbank Quarterly}, vol.~87, no.~2, pp.
  368--390, 2009.

\bibitem{kitchenham2004evidence}
B.~A. Kitchenham, T.~Dyba, and M.~Jorgensen, ``Evidence-based software
  engineering,'' in \emph{Proceedings. 26th International Conference on
  Software Engineering}.\hskip 1em plus 0.5em minus 0.4em\relax IEEE, 2004, pp.
  273--281.

\bibitem{kitchenham2009systematic}
B.~Kitchenham, O.~P. Brereton, D.~Budgen, M.~Turner, J.~Bailey, and S.~Linkman,
  ``Systematic literature reviews in software engineering--a systematic
  literature review,'' \emph{Information and software technology}, vol.~51,
  no.~1, pp. 7--15, 2009.

\bibitem{pizard2022longitudinal}
S.~Pizard, D.~Vallespir, and B.~Kitchenham, ``A longitudinal case study on the
  effects of an evidence-based software engineering training,'' in
  \emph{Proceedings of the ACM/IEEE 44th International Conference on Software
  Engineering: Software Engineering Education and Training}, 2022, pp. 1--13.

\bibitem{devanbu2016belief}
P.~Devanbu, T.~Zimmermann, and C.~Bird, ``Belief \& evidence in empirical
  software engineering,'' in \emph{Proceedings of the 38th international
  conference on software engineering}, 2016, pp. 108--119.

\bibitem{le2018bridging}
C.~Le~Goues, C.~Jaspan, I.~Ozkaya, M.~Shaw, and K.~T. Stolee, ``Bridging the
  gap: From research to practical advice,'' \emph{IEEE Software}, vol.~35,
  no.~5, pp. 50--57, 2018.

\bibitem{heroux2023research}
M.~A. Heroux, ``Research software science: Expanding the impact of research
  software engineering,'' \emph{Computing in Science \& Engineering}, 2023.

\bibitem{sims2022research}
B.~H. Sims, ``Research software engineering: Professionalization, roles, and
  identity,'' Los Alamos National Lab.(LANL), Los Alamos, NM (United States),
  Tech. Rep., 2022.

\bibitem{sufi2021rise}
S.~Sufi, ``The rise of a new digital third space professional in higher
  education: Recognising research software engineering,'' 2021.

\bibitem{mundt2021working}
M.~Mundt and R.~Milewicz, ``Working in harmony: Towards integrating rses into
  multi-disciplinary cse teams,'' \emph{Workshop on the Science of
  Scientific-Software Development and Use, sponsored by the U.S. Department of
  Energy, Office of Advanced Scientific Computing Research}, Dec 2021.

\bibitem{milewicz2021building}
R.~Milewicz and M.~Mundt, ``Building bridges: Establishing a dialogue between
  software engineering research and computational science,'' \emph{Workshop on
  the Science of Scientific-Software Development and Use, sponsored by the U.S.
  Department of Energy, Office of Advanced Scientific Computing Research}, Dec
  2021.

\bibitem{khangura2012evidence}
S.~Khangura, K.~Konnyu, R.~Cushman, J.~Grimshaw, and D.~Moher, ``Evidence
  summaries: the evolution of a rapid review approach,'' \emph{Systematic
  reviews}, vol.~1, no.~1, pp. 1--9, 2012.

\bibitem{cartaxo2019software}
B.~Cartaxo, G.~Pinto, B.~Fonseca, M.~Ribeiro, P.~Pinheiro, M.~T. Baldassarre,
  and S.~Soares, ``Software engineering research community viewpoints on rapid
  reviews,'' in \emph{2019 ACM/IEEE International Symposium on Empirical
  Software Engineering and Measurement (ESEM)}.\hskip 1em plus 0.5em minus
  0.4em\relax IEEE, 2019, pp. 1--12.

\bibitem{cartaxo2020rapid}
B.~Cartaxo, G.~Pinto, and S.~Soares, ``Rapid reviews in software engineering,''
  \emph{Contemporary Empirical Methods in Software Engineering}, pp. 357--384,
  2020.

\bibitem{niazi2015systematic}
M.~Niazi, ``Do systematic literature reviews outperform informal literature
  reviews in the software engineering domain? an initial case study,''
  \emph{Arabian Journal for Science and Engineering}, vol.~40, pp. 845--855,
  2015.

\bibitem{scurlock2014evidence}
L.~Scurlock-Evans, P.~Upton, and D.~Upton, ``Evidence-based practice in
  physiotherapy: a systematic review of barriers, enablers and interventions,''
  \emph{Physiotherapy}, vol. 100, no.~3, pp. 208--219, 2014.

\bibitem{kasoju2013analyzing}
A.~Kasoju, K.~Petersen, and M.~V. M{\"a}ntyl{\"a}, ``Analyzing an automotive
  testing process with evidence-based software engineering,'' \emph{Information
  and Software Technology}, vol.~55, no.~7, pp. 1237--1259, 2013.

\bibitem{pizard2023assessing}
S.~Pizard, F.~Acerenza, D.~Vallespir, and B.~Kitchenham, ``Assessing attitudes
  towards evidence-based software engineering in a government agency,''
  \emph{Information and Software Technology}, vol. 154, p. 107101, 2023.

\bibitem{holton2007coding}
J.~A. Holton, ``The coding process and its challenges,'' \emph{The Sage
  handbook of grounded theory}, vol.~3, pp. 265--289, 2007.

\bibitem{EBPDataset}
\BIBentryALTinterwordspacing
R.~M. Milewicz, J.~Bisila, M.~Mundt, and J.~Teves, ``{Rapid Review Dataset for
  Seeking Enlightenment: Incorporating Evidence-Based Practice Techniques in a
  Research Software Engineering Team},'' Jul. 2023. [Online]. Available:
  \url{https://doi.org/10.5281/zenodo.8169610}
\BIBentrySTDinterwordspacing

\bibitem{freund2015contextualizing}
L.~Freund, ``Contextualizing the information-seeking behavior of software
  engineers,'' \emph{Journal of the Association for Information Science and
  Technology}, vol.~66, no.~8, pp. 1594--1605, 2015.

\bibitem{hartling2015taxonomy}
L.~Hartling, J.-M. Guise, E.~Kato, J.~Anderson, S.~Belinson, E.~Berliner, D.~M.
  Dryden, R.~Featherstone, M.~D. Mitchell, M.~Motu'Apuaka \emph{et~al.}, ``A
  taxonomy of rapid reviews links report types and methods to specific
  decision-making contexts,'' \emph{Journal of Clinical Epidemiology}, vol.~68,
  no.~12, pp. 1451--1462, 2015.

\bibitem{borah2017analysis}
R.~Borah, A.~W. Brown, P.~L. Capers, and K.~A. Kaiser, ``Analysis of the time
  and workers needed to conduct systematic reviews of medical interventions
  using data from the prospero registry,'' \emph{BMJ open}, vol.~7, no.~2, p.
  e012545, 2017.

\bibitem{rousseau2016evidence}
D.~M. Rousseau and B.~C. Gunia, ``Evidence-based practice: The psychology of
  ebp implementation,'' \emph{Annual review of psychology}, vol.~67, pp.
  667--692, 2016.

\bibitem{melnyk2004nurses}
B.~M. Melnyk, E.~Fineout-Overholt, N.~Fischbeck~Feinstein, H.~Li, L.~Small,
  L.~Wilcox, and R.~Kraus, ``Nurses' perceived knowledge, beliefs, skills, and
  needs regarding evidence-based practice: Implications for accelerating the
  paradigm shift,'' \emph{Worldviews on Evidence-Based Nursing}, vol.~1, no.~3,
  pp. 185--193, 2004.

\bibitem{kiss2010self}
T.~L. Kiss, M.~O'Malley, and T.~J. Hendrix, ``Self-efficacy-based training for
  research literature appraisal: a competency for evidence-based practice,''
  \emph{Journal for Nurses in Professional Development}, vol.~26, no.~4, pp.
  170--177, 2010.

\bibitem{leeman2017developing}
J.~Leeman, L.~Calancie, M.~C. Kegler, C.~T. Escoffery, A.~K. Herrmann,
  E.~Thatcher, M.~A. Hartman, and M.~E. Fernandez, ``Developing theory to guide
  building practitioners’ capacity to implement evidence-based
  interventions,'' \emph{Health Education \& Behavior}, vol.~44, no.~1, pp.
  59--69, 2017.

\bibitem{belden2012effect}
C.~V. Belden, J.~Leafman, G.~Nehrenz, and P.~Miller, ``The effect of evidence
  based practice on workplace empowerment of rural registered nurses,''
  \emph{Online Journal of Rural Nursing and Health Care}, vol.~12, no.~2, pp.
  64--76, 2012.

\bibitem{dalheim2012factors}
A.~Dalheim, S.~Harthug, R.~M. Nilsen, and M.~W. Nortvedt, ``Factors influencing
  the development of evidence-based practice among nurses: a self-report
  survey,'' \emph{BMC health services research}, vol.~12, pp. 1--10, 2012.

\bibitem{melnyk2012state}
B.~M. Melnyk, E.~Fineout-Overholt, L.~Gallagher-Ford, and L.~Kaplan, ``The
  state of evidence-based practice in us nurses: critical implications for
  nurse leaders and educators,'' \emph{JONA: The Journal of Nursing
  Administration}, vol.~42, no.~9, pp. 410--417, 2012.

\bibitem{wohlin2021challenges}
C.~Wohlin and A.~Rainer, ``Challenges and recommendations to publishing and
  using credible evidence in software engineering,'' \emph{Information and
  software technology}, vol. 134, p. 106555, 2021.

\bibitem{callaham2015expert}
M.~L. Callaham, ``Expert opinion: supplementing the gaps in evidence-based
  medicine,'' \emph{Annals of Emergency Medicine}, vol.~65, no.~1, pp. 61--62,
  2015.

\bibitem{klein2016can}
D.~E. Klein, D.~D. Woods, G.~Klein, and S.~J. Perry, ``Can we trust best
  practices? six cognitive challenges of evidence-based approaches,''
  \emph{Journal of Cognitive Engineering and Decision Making}, vol.~10, no.~3,
  pp. 244--254, 2016.

\bibitem{goldstein2018practice}
K.~M. Goldstein, D.~Vogt, A.~Hamilton, S.~M. Frayne, J.~Gierisch, J.~Blakeney,
  A.~Sadler, B.~M. Bean-Mayberry, D.~Carney, B.~DiLeone \emph{et~al.},
  ``Practice-based research networks add value to evidence-based quality
  improvement,'' in \emph{Healthcare}, vol.~6, no.~2.\hskip 1em plus 0.5em
  minus 0.4em\relax Elsevier, 2018, pp. 128--134.

\bibitem{neilson2015moving}
E.~Neilson, K.~C. Smith, D.~Steinwachs, D.~F. Phelan-Emrick, R.~S. Lawrence,
  J.~V. Bowie, and B.~Cohen, ``Moving research into practice: the diffusion of
  evidence-based recommendations through professional societies,'' in
  \emph{Implementation Science}, vol.~10, no.~1.\hskip 1em plus 0.5em minus
  0.4em\relax BioMed Central, 2015, pp. 1--2.

\bibitem{rodwin2001commentary}
M.~A. Rodwin, ``Commentary: The politics of evidence-based medicine,''
  \emph{Journal of Health Politics, Policy and Law}, vol.~26, no.~2, pp.
  439--446, 2001.

\bibitem{gambrill2018evidence}
E.~Gambrill, ``Evidence-based practice: An alternative to authority-based
  practice (revisiting our heritage),'' \emph{Families in Society}, vol.~99,
  no.~3, pp. 283--294, 2018.

\bibitem{jacobs2015principles}
J.~P. Jacobs, G.~Wernovsky, D.~S. Cooper, and T.~R. Karl, ``Principles of
  shared decision-making within teams,'' \emph{Cardiology in the Young},
  vol.~25, no.~8, pp. 1631--1636, 2015.

\bibitem{kutash2014quality}
K.~Kutash, M.~Acri, M.~Pollock, K.~Armusewicz, S.-c. Serene~Olin, and K.~E.
  Hoagwood, ``Quality indicators for multidisciplinary team functioning in
  community-based children’s mental health services,'' \emph{Administration
  and Policy in Mental Health and Mental Health Services Research}, vol.~41,
  pp. 55--68, 2014.

\bibitem{mathieson2019strategies}
A.~Mathieson, G.~Grande, and K.~Luker, ``Strategies, facilitators and barriers
  to implementation of evidence-based practice in community nursing: a
  systematic mixed-studies review and qualitative synthesis,'' \emph{Primary
  health care research \& development}, vol.~20, p.~e6, 2019.

\bibitem{kelly2007software}
D.~F. Kelly, ``A software chasm: Software engineering and scientific
  computing,'' \emph{IEEE software}, vol.~24, no.~6, pp. 120--119, 2007.

\end{thebibliography}

\end{document}